\documentclass[aps,prl,twocolumn,
superscriptaddress]{revtex4-1}
\usepackage{amsmath}
\usepackage{graphicx}
\usepackage{amsfonts}
\usepackage{amssymb}
\usepackage{color}

\newcommand{\up}{\uparrow}
\newcommand{\down}{\downarrow}
\renewcommand{\k}{{\bf k}}
\newcommand{\p}{{\bf p}}
\newcommand{\q}{{\bf q}}

\newcommand{\0}{{\bf 0}}

\newcommand{\ef}{\epsilon_F}
\newcommand{\eq}{\epsilon_{\q}}

\newcommand{\nn}{\nonumber}

\newcommand{\D}{{\cal D}}
\begin{document}

\title{The $p$-wave polaron}

\author{Jesper Levinsen}
\email{jfl36@cam.ac.uk}
\affiliation{T.C.M. Group, Cavendish Laboratory, JJ Thomson Avenue, Cambridge CB3 0HE, United Kingdom}

\author{Pietro Massignan}
\affiliation{ICFO - The Institute of Photonic Sciences - 08860 Castelldefels (Barcelona), Spain}

\author{Fr\'ed\'eric Chevy}
\affiliation{Laboratoire Kastler Brossel - ENS, Universit\'e Paris 6, CNRS, 24 rue Lhomond, 75005 Paris, France}

\author{Carlos Lobo}
\affiliation{School of Mathematics, University of Southampton, Highfield, Southampton, SO17 1BJ, United Kingdom}

\date{\today}

\begin{abstract}
  We consider the properties of a single impurity immersed in a Fermi
  sea close to an interspecies $p$-wave Feshbach resonance. We
  calculate its dispersion and spectral response to a radiofrequency
  pulse. In the presence of a magnetic field, dipolar interactions
  split the resonance and lead to the appearance of two novel features
  with respect to the $s$-wave case: a third polaron branch in the
  excitation spectrum, in addition to the usual attractive and
  repulsive ones; and an anisotropic dispersion of the impurity
  characterized by different effective masses perpendicular and
  parallel to the magnetic field. The anisotropy can be tuned as a
  function of the field strength and the two effective masses may have
  opposite signs, or become smaller than the bare mass.
\end{abstract}


\maketitle

Understanding the physics of a single impurity in a degenerate
ultracold gas has been essential in discovering the phase diagram of
spin-imbalanced Fermi mixtures \cite{Lobo2006,Chevy2006,Pilati2008}.
Close to a Feshbach resonance, the impurity state becomes truly
many-body in character. A large effort has been devoted to its
understanding, resulting in an impressive agreement between theory and
experiment in recent years \cite{ChevyMora2010}. The emergent picture
is that the quasiparticle formed by the impurity interacting with the
background gas can be fermionic (a ``polaron") or bosonic (a
``molecule").

So far, the polaron has not been studied close to higher partial wave
resonances. Particularly interesting are $p$-wave resonances, as
$p$-wave coupled superfluids are predicted to display a richer phase
diagram than their $s$-wave counterparts
\cite{Gurarie2005a,Cheng2005,Gurarie2006,Iskin2006}. For example, the
Bardeen-Cooper-Schrieffer (BCS) and Bose-Einstein condensation (BEC)
regimes of these superfluids are separate phases \cite{Volovik2003},
and each of these can be either chiral or polar
\cite{Gurarie2006}. When confined to two dimensions, the superfluid
BCS phase is even topologically nontrivial \cite{Read2000}. Therefore
it becomes crucial to understand the nature of the quasiparticles
which could form such states. For $^6$Li and $^{40}$K, $p$-wave
resonances between atoms in the same hyperfine state were found to
have an extremely narrow magnetic width, of order $\lesssim 1G$
\cite{Zhang2004,Ticknor2004}.  However, $p$-wave resonances in Li-K
mixtures \cite{Wille2008} proved to have larger magnetic widths, in
the range $1-10$G, indicating a stronger open-channel character of the
$p$-wave interaction. Given the stability of magnetic fields in state
of the art experiments ($\approx\pm 1$mG), such resonances are now
finally accessible for detailed study.

\begin{figure}[ht]
\includegraphics[width=\linewidth]{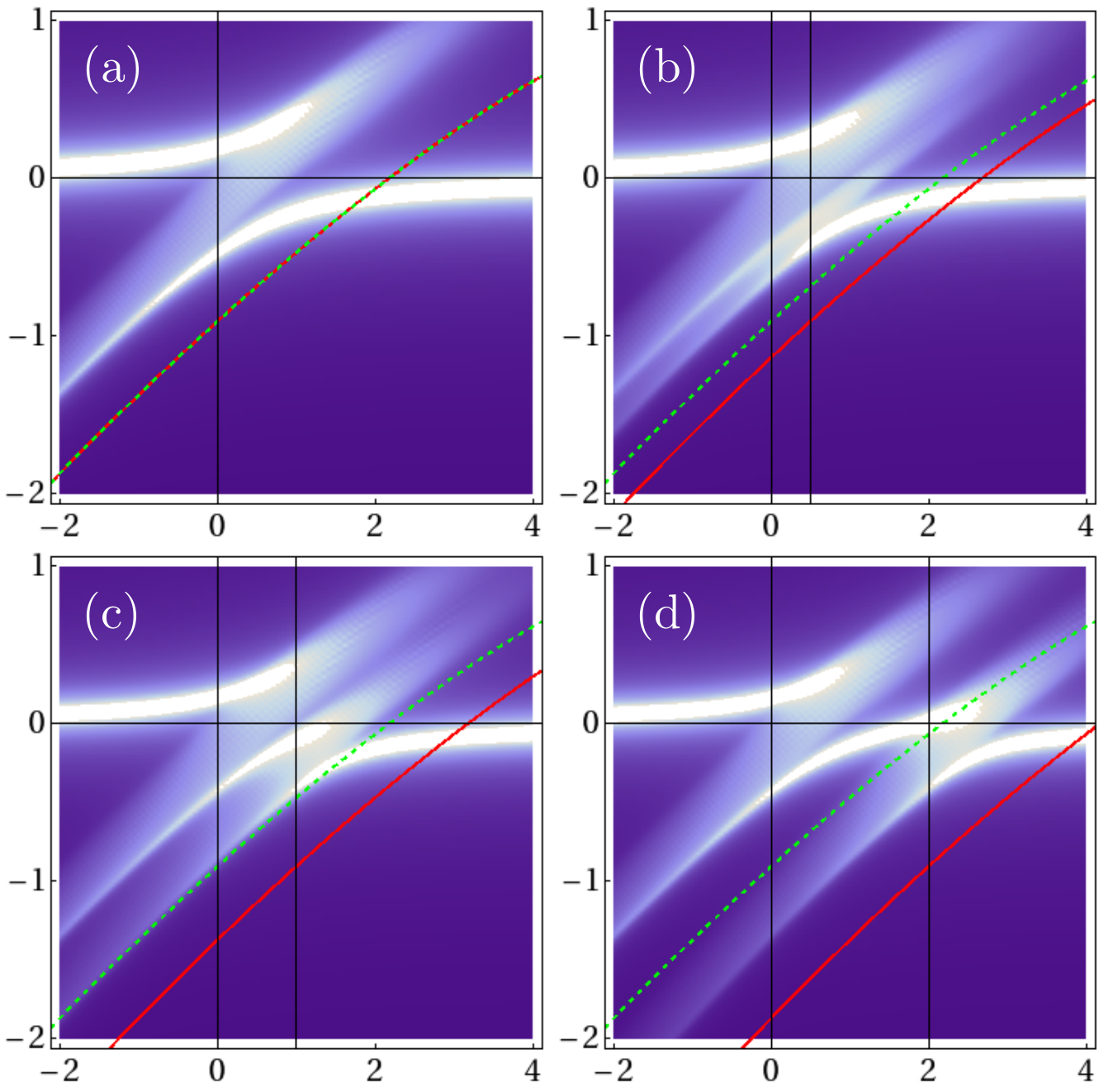}
\caption{(color online). Spectral function of the $p$-wave
  polaron. The horizontal and vertical axes are $-(k_F^3v_{\pm
    1})^{-1}$ and $\omega/\epsilon_F$. The $m_l=\pm 1$ and $m_l=0$
  resonances are located respectively at $x=0$ and $x=\delta$, with
  $\delta=$0(a), 0.5(b), 1(c), 2(d). Thick lines are the dressed
  molecules with $m_l=0$ (solid) and $m_l=\pm1$ (dashed).  Here
  $-k_0/k_F=10$.}
\label{spectralFunctionSplit}
\end{figure}

Here we study the $p$-wave polaron problem by considering a single
impurity atom, labelled $\down$, immersed in a spin-$\up$ Fermi sea,
assuming that the two atomic species have the same mass and are
strongly interacting in the $p$-wave channel, as may be achieved close
to a $p$-wave Feshbach resonance. We will neglect any background
$s$-wave interactions (see below).

An interesting feature of $p$-wave Feshbach resonances is that they
are generally split into a doublet according to the projection $m_l$
of the relative angular momentum onto the magnetic field axis. As
first discussed in Ref. \cite{Ticknor2004}, the $m_l=\pm 1$ resonances
are shifted towards higher energy from the $m_l=0$ by the magnetic
dipole-dipole interaction between the outer shell electrons in the
presence of the external magnetic field. As we shall see, an important
consequence is that when the energy splitting between the two
molecular levels is larger than the width of the $m_l=0$ molecule-hole
continuum, a new polaron branch appears lying between the attractive
and repulsive polaron branches which had been previously observed
close to $s$-wave Feshbach resonances
\cite{Kohstall2011,Koschorreck2012}. This is reflected in the spectral
function of the impurity atom displayed in
Fig. \ref{spectralFunctionSplit}. Additionally, we find a regime close
to the $m_l=0$ resonance in which the effective mass of the attractive
polaron becomes negative (positive) along (perpendicular to) the
magnetic field axis with the opposite behaviour near the $m_l=\pm 1$
resonance. Clearly, these two new features could lead to many-body
behaviour (e.g. collective modes, possible superfluid states) which
would be dramatically different from that of the $s$-wave system.

The impurity may also bind to a particle from the Fermi sea to form a
diatomic molecule, and as the interactions are increased this
molecular state becomes energetically favorable. We calculate the
energies of the polaron and the molecule, and show that the position
of the polaron-molecule transition shifts towards the BCS side for
increasingly narrow resonances.

To model the $p$-wave Feshbach resonance we use the following
two-channel Hamiltonian \cite{Chevy2005,Gurarie2006}, and work in
units where $\hbar=1$:
\begin{eqnarray}
&&\hspace{-5mm}
  H=\sum_{\p,\sigma=\up,\down}\frac{p^2}{2m} a^\dag_{\sigma{\p}}
  a_{\sigma{\p}}+\sum_{{\q},\mu}\left(\epsilon_\mu+
    \frac{q^2}{4m}\right) b^\dag_{\mu\q}b_{\mu\q} \nn
\\ &&\hspace{-6mm}
  +\sum_{\p,\q,\mu}\frac{g(|\p|)}{\sqrt V}\p_\mu \left(b^\dag_{{\bf q}\mu}
    a_{\up\frac{\q}{2}+\p}a_{\down\frac{\q}{2}-\p}+
    a^\dag_{\down\frac{\q}{2}-\p}a^\dag_{\up\frac{\q}{2}+\p}b_{{\bf q}\mu}
\right).
\label{eq:H}
\end{eqnarray}
For convenience we define bosonic operators $b_{\mu=x,y,z}$ in terms
of the closed channel $l=1$ molecule operators $b_{m_l=0,\pm1}$ as
$b_x \equiv (b_{+1}+b_{-1})/\sqrt{2}$, $b_y \equiv
-i(b_{+1}-b_{-1})/\sqrt{2}$, $b_z \equiv b_0$.  We specialize to the
case where the $m_l=\pm1$ resonances are
degenerate~\footnote{Otherwise time-reversal symmetry is broken and an
  additional term appears in the
  Hamiltonian~\cite{Gurarie2006}}. $a_\sigma$ and $a_\sigma^\dagger$
are the creation and annihilation operators of fermions with mass $m$,
and $V$ is the system volume. The resonance splitting caused by
dipolar anisotropy may be modelled by a positive shift of the
$m_l=\pm1$ molecule energies: $\epsilon_{x,y}=\epsilon_z+\delta_0$
with $\epsilon_{x,y}=\epsilon_{\pm1}$ and $\epsilon_z=\epsilon_0$.

Interactions couple the closed channel molecule to a pair of atoms in
the open channel, and are described by the momentum-dependent coupling
constant $g(|\p|)\p_\mu$.  The coupling vanishes above a cutoff
$\Lambda$ of the order of $1/R_e$, the inverse van der Waals
length. In the low energy limit, the physical results should not
depend on the actual shape of $g$ and we choose the cutoff function to
be proportional to a step function, $g(|\p|)=g\Theta(\Lambda-p)$.

\begin{figure}
\centering
\includegraphics[width=\linewidth]{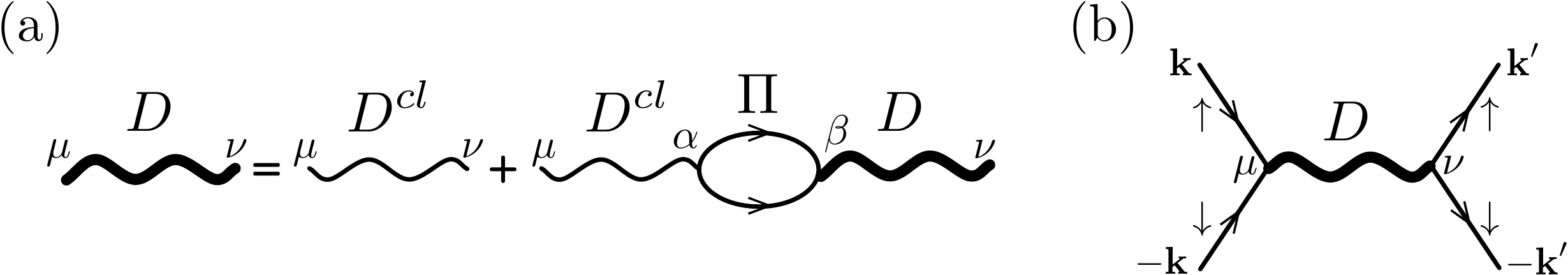}
\caption{(a) The renormalized propagator of molecules (thick wavy
  line). The thin wavy line is the bare molecule while the straight
  lines are fermions. (b) The diagram which leads to the $T$-matrix,
  Eq. (\ref{eq:tkk}).}
\label{fig:1}
\end{figure}

In the absence of the Fermi sea, the bare propagator of the closed
channel molecule can be read off from the Hamiltonian as
\begin{equation}
D^{cl}_{\mu\nu}(\p,\omega)=\frac{\delta_{\mu\nu}}{\omega-
\epsilon_\mu-p^2/4m+i0}
\equiv D^{cl}_\mu(\p,\omega)\delta_{\mu\nu}.
\end{equation}
The renormalized molecular propagator $D^0_{\mu\nu}$ is dressed by the
polarization bubble $\Pi^0$ as shown in Fig. \ref{fig:1}a. In vacuum
$\Pi^0_{\mu\nu}\equiv \Pi^0\delta_{\mu\nu}$ is diagonal and takes the
form
\begin{equation}
\Pi^0(\p,\omega)=\frac13\int\frac{d^3q}{(2\pi)^3}\frac{q^2g^2(q)}
{\omega-q^2/m-p^2/4m+i0}.
\end{equation}
Thus the propagator in vacuum is also diagonal,
i.e. $D^0_{\mu\nu}(\p,\omega) =\delta_{\mu\nu}/
\left\{[D^{cl}_\mu(\p,\omega)]^{-1}-\Pi^0(\p,\omega)\right\}$.

The elastic scattering between a spin-$\up$ and a spin-$\down$ atom in
the ladder approximation is described by a $T$-matrix. The vacuum
$T$-matrix in the present problem is given by (see Fig. \ref{fig:1}b)
\begin{equation}
  T(\k,\k')=\sum_\mu
  \frac{g^2(k) k_\mu k_\mu'}{[D^{cl}_\mu ({\bf 0},
    k^2/m)]^{-1}-\Pi^0 ({\bf 0},k^2/m)}.
\label{eq:tkk}
\end{equation}
Note that if the detuning $\epsilon_\mu$ is independent of $\mu$ the
interaction is isotropic and the $T$-matrix is proportional to
$\k\cdot\k'$. However, the splitting of the $p$-wave resonance by the
dipole-dipole interaction in the presence of a magnetic field breaks
rotational symmetry. In order to relate the parameters of the model to
the physical observables, we compare with the low energy expansion of
the $p$-wave scattering amplitude $f(\k,\k')\equiv-(m/4\pi)T(\k,\k')$:
\begin{equation}
f(\k,\k')\approx\sum_\mu \frac{k_\mu k_\mu'}{-v_\mu^{-1}+\frac12k_0k^2-ik^3}.
\end{equation}
Here $v_\mu$ is the state dependent scattering volume and $k_0$, with
dimension of momentum, is the $p$-wave analogue of the effective range
\footnote{In general, $k_0$ will also depend on $m_l$ through the
  choice of cut-off function, however for simplicity we assume $k_0$
  to be a constant}. Evaluating the polarization bubble and inserting
in Eq.\ (\ref{eq:tkk}) yields the relationships
\begin{equation}
v_\mu = -\frac{mg^2}{12\pi\left(\epsilon_\mu-
\frac{mg^2\Lambda^3}{18\pi^2}\right)}, \hspace{5mm}
k_0=-\frac{24\pi}{m^2g^2}(1+c_2),
\label{pWaveCoeffs}
\end{equation}
with $c_2\equiv m^2g^2\Lambda/6\pi^2$.  The vacuum molecule propagator
is then
\begin{eqnarray}
\hspace{-5mm}D^0_{\mu\nu}(\p,\omega)= & & \nn \\ && \hspace{-14mm}
\frac{-12\pi/(mg^2)\delta_{\mu\nu}}
{-v_\mu^{-1}+\frac12k_0(m\omega-p^2/4)-(p^2/4-m\omega-i0)^{3/2}}.
\end{eqnarray}
Note that the parameter $k_0$ is naturally large and negative, of the
order of the cut-off $\Lambda$ \cite{Levinsen2007,Jona-Lasinio2008}.

The $p$-wave resonances in the many-body system are characterized by
two dimensionless parameters. Of these $\gamma\sim (1+c_2)k_F/k_0$
controls interactions at the scale of the Fermi momentum $k_F$
(i.e. many-body physics). Resonances with $\gamma\ll1$ ($\gamma\gg1$)
are termed narrow (wide) and quantum fluctuations are suppressed for
$\gamma\ll1$ \cite{Gurarie2006}. The second parameter $c_2$ controls
interactions at the scale $\Lambda$ (i.e. few-body physics)
\cite{Levinsen2007} and distinguishes strongly coupled ($c_2\gg1$)
from weakly coupled $(c_2\ll1)$ systems. Identical fermions
interacting close to a strong Feshbach resonance may form trimer
states \cite{Jona-Lasinio2008}. However, using the method of
Ref. \cite{Levinsen2007}, we have checked that trimers consisting of
two identical fermions and an impurity of equal mass do not form in
vacuum in the absence of resonance splitting.

\begin{figure}
\centering
\includegraphics[width=0.6\linewidth]{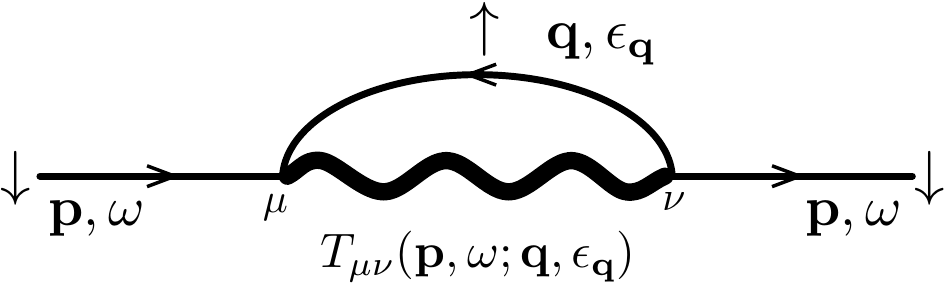}
\caption{The polaron self energy.}
\label{polaronSelfEnergy}
\end{figure}

We now turn to the question of the many-body state of a spin-$\down$
impurity immersed in a spin-$\up$ Fermi sea at zero temperature. In
the $s$-wave case both the attractive and repulsive polaron branches
are accurately described by dressing the impurity with a single
particle-hole excitation \cite{Chevy2006,Combescot2007a}; we will
adopt the same approximation for the $p$-wave case. Indeed, this
approximation becomes exact far away from resonance, or in the limit
of narrow resonances, $\gamma\ll1$ \cite{Gurarie2006}. The propagator
of the impurity with momentum $\p$ and energy $\omega$ in the medium
is
\begin{equation}
G_\down(\p,\omega)=\frac1{\omega-p^2/2m-\Sigma(\p,\omega)+i0},
\end{equation}
with $\Sigma$ the self-energy. The energy of the polaron satisfies
$E=\mbox{Re}[\Sigma(\p,E)]$.  The self energy in the single
particle-hole approximation is given by the diagram in
Fig. \ref{polaronSelfEnergy}:
\begin{equation}
\Sigma(\p,\omega)=\int\frac{d^3\q}{(2\pi)^3}n_{F\up}(\q)T(\p,\omega;\q,\eq),
\label{eq:Sigma}
\end{equation}
where the Fermi function $n_{F\up}(\q)$ takes the value 1 if the state
with momentum $\q$ is occupied, 0 otherwise. The off-shell
two-particle scattering $T$-matrix in the medium is
\begin{eqnarray}
 T(\p,\omega;\q,\eq) & = & g^2(|\p-\q|/2)\sum_{\mu\nu}\left(\frac{\p-\q}2\right)_\mu
  \left(\frac{\p-\q}2\right)_\nu \nn \\ && \hspace{-25mm}
\times\left\{\left[D^{cl}(\p+\q,\omega+\eq)\right]^{-1}-\Pi(\p+\q,\omega+
    \eq)\right\}^{-1}_{\mu\nu}.
\label{eq:tmed}
\end{eqnarray}
Here $(\p,\omega)$ [$(\q,\eq)$] are the $\up$ [$\down$] atom momentum
and energy entering the $T$-matrix, with $\eq=q^2/2m$.  The
polarization bubble in the medium is given by
\begin{equation}
  \Pi_{\mu\nu}(\q,\omega)=\int\frac{d^3 \k}{(2\pi)^3}\frac{k_\mu k_\nu
    g^2(k)\left[1-n_{F\up}(\k+\q/2)\right]}{\omega-q^2/4m-k^2/m+i0}.
\label{eq:pimed}
\end{equation}
Since $|\q|\ll \Lambda$ it is useful to write
$\Pi=\Pi^0+(\Pi-\Pi^0)$. Then, upon proper renormalization, the
coupling in Eqs. (\ref{eq:tmed}) and (\ref{eq:pimed}) reduces to the
bare value.

The polarization bubble $\Pi$ is a tensor, as collisions with
particles in the Fermi sea generally do not preserve the projection of
the molecule's angular momentum. However, $D^{cl}$ is diagonal, and
the matrix inverse in Eq. (\ref{eq:tmed}) may be performed as follows.
We write the polarization bubble as
\begin{equation}
\Pi_{\mu\nu}(\p+\q,\omega+\eq)\equiv A\delta_{\mu\nu}
+B (\p+\q)_\mu (\p+\q)_\nu/|\p+\q|^2.
\end{equation}
In the matrix inverse of Eq. (\ref{eq:tmed}), sums containing only $A$
or only $B$ terms form geometric series. Resumming first the $A$
terms, and successively adding the $B$ terms, one finds
\begin{equation}
T(\p,\omega;\q,\eq)=
\frac{g^2}{4}
\left[\D_{-}+\frac{\D_{\rm x}^2}{|\p+\q|^2/B-\D_{+}}\right],
\label{mediumTmatrix}
\end{equation}
with $\D_{\pm}\equiv\sum_\mu\frac{(\p\pm\q)_\mu^2}
{[D_{\mu}^{cl}(\p+\q,\omega+\eq)]^{-1}-A}$ and $\D_{\rm
  x}\equiv\sum_\mu\frac{(\p+\q)_\mu (\p-\q)_\mu}
{[D_{\mu}^{cl}(\p+\q,\omega+\eq)]^{-1}-A}$.

The above formalism allows us to calculate the spectral function of
the impurity atoms, defined as
$A_\down(\p,\omega)=-2\mbox{Im}[G_\down(\p,\omega)]$, which gives the
spectral response of the impurity to a radiofrequency pulse of
frequency $\omega$. In particular, the spectral function peaks at the
energy of the quasiparticle states with a finite wavefunction overlap
with the bare impurity, {\em i.e.} the polarons. Our results for the
spectral function close to resonance are shown in
Fig. \ref{spectralFunctionSplit} for several values of the
dimensionless resonance splitting $\delta\equiv\delta_0
12\pi/(mg^2k_F^3)$. In the absence of resonance splitting, the picture
is qualitatively the same as in the $s$-wave polaron case
\cite{Chevy2006,Cui2010,Massignan:2011fk} with an attractive
(repulsive) polaron of energy lower (higher) than the bare
impurity. However, an additional branch appears for finite resonance
splitting. For small resonance splitting,
$\delta_0\lesssim\epsilon_F$, the intermediate branch disappears in
the $m_l=0$ molecule-hole continuum. For larger resonance splitting
the intermediate branch shows up as an isolated spectral line which is
a repulsive polaron close to the $m_l=0$ resonance, approximately a
free impurity between the resonances, and an attractive polaron close
to the $m_l=\pm1$ resonance.

\begin{figure}
\includegraphics[width=\linewidth]{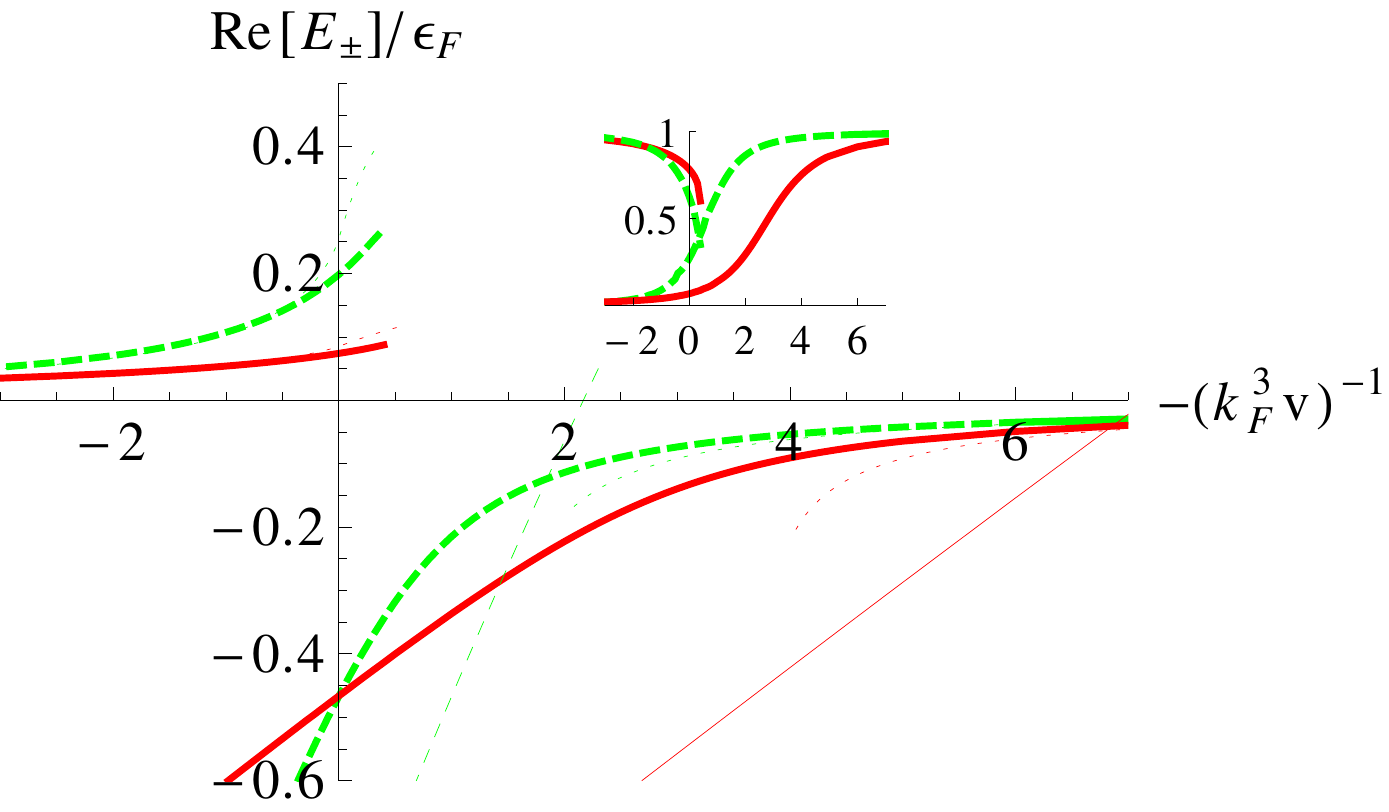}
\caption{(color online). Energy (main) and residue (inset) of the two
  $p$-wave polarons for $\delta=0$ through the crossover for range
  parameters: $-k_0/k_F=$30(solid), 10(dashed). The dotted lines are
  the perturbative result, Eq.  (\ref{polaronEnergy2ndOrder}), and the
  thin lines are the molecule energies as given by the Thouless pole.}
\label{twoPolaronsThroughTheCrossover}
\end{figure}

In addition to energy, the polaron is characterized by its
quasiparticle weight, the residue, given by $Z=\left[1-\partial
  \mbox{Re}\left\{\Sigma(\0,E)\right\}/\partial E\right]^{-1}$
evaluated at the quasiparticle energy.  In
Fig. \ref{twoPolaronsThroughTheCrossover} we display the energy and
residue across resonance for various values of $k_0/k_F$. We see that
the absolute values of the polaron energies decrease for increasing
$|k_0/k_F|$ as expected since particle-hole fluctuations become
suppressed \cite{Gurarie2006}. Indeed, in this limit the self energy
(\ref{eq:Sigma}), in the absence of resonance splitting ($v_\mu\equiv
v$), takes the form
\begin{equation}
\Sigma(\0,E)=-\frac{3}{2\pi m}\int_0^{k_F}\frac{q^4dq}{-v^{-1}+
\frac12mk_0(E+\eq/2)+i0}.
\label{polaronEnergy2ndOrder}
\end{equation}

As mentioned above, the impurity may also bind a particle from the
majority Fermi sea to form a bosonic quasiparticle --- a molecule. In
the ``Cooper pair" approximation (i.e., with no particle-hole pairs),
the energy $E_{\rm mol}$ of the molecular state is given by the pole
of the $T$-matrix at $\p=0$ and $\omega=E_{\rm mol}+\epsilon_F$.  The
latter approximation is an upper bound to the real energy of the
molecule, which becomes exact for $|k_0/k_F|\gg1$.  On the BEC side,
$E_{\rm mol}$ tends to $E_b-\epsilon_F$, with $E_b$ the energy of the
molecule in vacuum. The molecule energy is included in
Figs. \ref{spectralFunctionSplit} and
\ref{twoPolaronsThroughTheCrossover}. The latter illustrates that as
the resonance becomes more narrow (for increasing $-k_0/k_F$) the
transition from a polaronic to a molecular ground state takes place
further towards the BCS limit, as is the case close to a narrow
$s$-wave Feshbach resonance
\cite{Massignan2011,Qi2011,Trefzger2011,Kohstall2011}. Note, however,
that the limit of narrow $p$-wave resonances corresponds to small
densities, while the opposite is true for $s$-wave resonances.

\begin{figure}
\centering
\includegraphics[width=\linewidth]{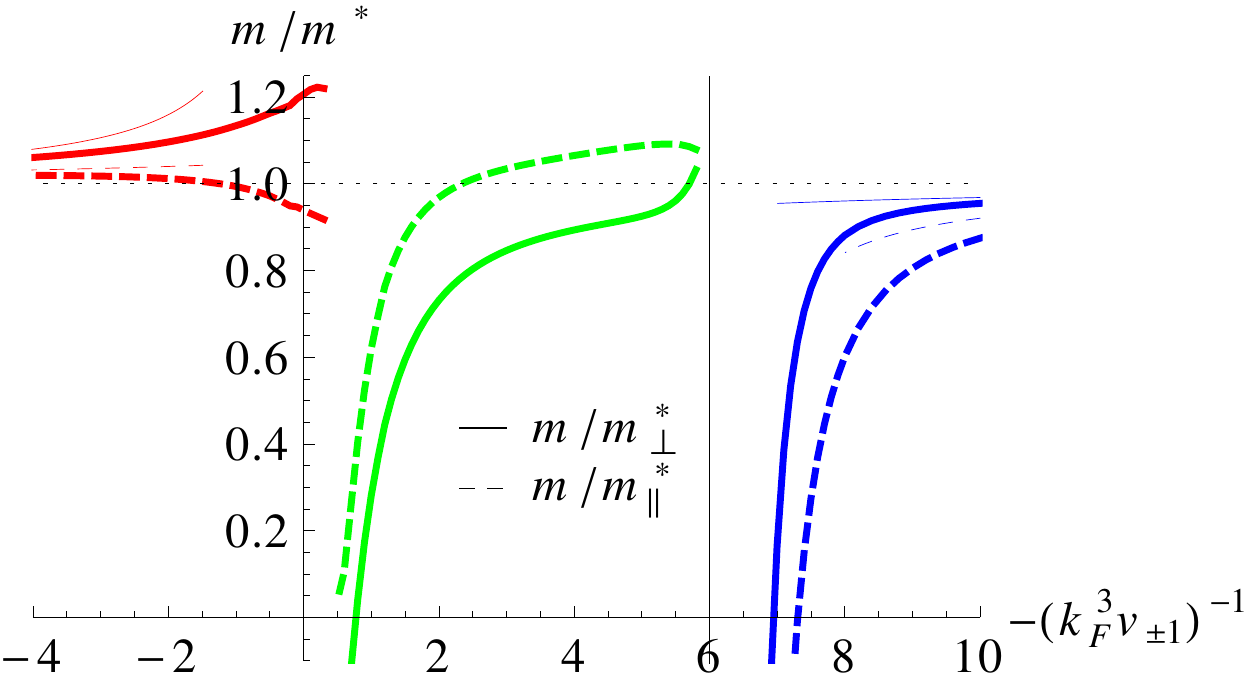}
\caption{(color online). Inverse effective mass of $p$-wave polarons
  moving perpendicular (solid) and parallel (dashed) to the magnetic
  field. Here $-k_0/k_F=$10, and the two resonances are split by
  $\delta=6$, i.e., the $m_l=\pm 1$ ($m_l=0$) resonance lies at $x=0$
  ($x=6$).  The thin lines are the analytic weak-coupling results
  $m/m_{\perp, \parallel}^*\approx 1+v_{\perp, \parallel} k_F^3/\pi$,
  valid in the regime where $|v_{\perp, \parallel} k_F^3|^{-1} \gg
  |k_0/k_F|$.}
\label{mEffSplit}
\end{figure}

The presence of a magnetic field which splits the resonances also
breaks isotropy since it differentiates between $m_l$ levels. One
consequence is the appearance of a strongly anisotropic dispersion of
the impurity close to either of the resonances.  At small momenta, the
dispersion may be written as
$E(\p)=E(\0)+p_\parallel^2/2m_{\parallel}^*+p_\perp^2/2m_{\perp}^*$,
where $p_\parallel$ and $p_\perp$ are the projections of $\p$ along
the magnetic field, and on the plane perpendicular to it.  We compute
the effective mass tensor, showing how, on the BEC side, its magnitude
is generally smaller than the bare mass of the particle. This is
allowed since, in this region, the polaron is no longer the ground
state. Moreover, close to the $m_l=0$ resonance $m/m^*_\parallel$
becomes negative while $m/m^*_\perp$ is still positive, the opposite
behavior occurring in the vicinity of the $m_l=\pm1$ resonance. This
is illustrated in Fig.\ \ref{mEffSplit}.

We now make a few final remarks. In the above, any background $s$-wave
scattering between the two atomic species has been ignored. This is
justified provided the background scattering is small, in particular
for a $p$-wave resonance close to a zero-crossing of an $s$-wave
resonance. At vanishing scattering energy the $s$-wave channel
dominates.  However, in the present system the low energy scale is set
by $\ef$. Then the total cross section for scattering of two atoms in
the $s$-wave channel is $\sigma_s=4\pi a^2$ while the $p$-wave cross
section is $\sigma_p=12\pi|f_p(k_F)|^2$ \cite{LL}. If
$|vk_F^3|^{-1}\gg |k_0/k_F|$ then $p$-wave scattering is dominant for
$|vk_F^3|\gtrsim |k_Fa|$ while in the opposite case we require
$|k_0a|\lesssim1$. Both requirements are achievable provided $k_0$ is
not too large.

Early experimental studies of $p$-wave Feshbach molecules showed that
these were short-lived, with lifetimes in the range 2 to 20 ms
\cite{Zhang2004,Gaebler2007,Fuchs2008,Inada2008}. This was explained
in Refs. \cite{Levinsen2007,Jona-Lasinio2008,Levinsen2008} as due to
relaxation processes as well as possible recombination into trimers.
However, radiofrequency spectroscopy has been able to study in detail
the complete spectral response of metastable many-body states of
strongly-interacting fermions with an intrinsic lifetime of only 30ms
\cite{Kohstall2011} including their polaronic and molecular
branches. Furthermore, a recent proposal to modify the $p$-wave
resonant interaction by coupling to a long-range excited state with
the use of an optical Feshbach resonance predicts the possibility of
suppressing three-body recombination \cite{Goyal2010}, potentially
allowing for longer lifetimes.

In this Letter, we studied the properties of the $p$-wave
polaron. These are observable by radiofrequency spectroscopy. In
particular, the anisotropic dispersion relation may be studied using
angle resolved photo emission spectroscopy. Our results are also
relevant to studies of resonantly enhanced atom-dimer scattering near
the appearance of confinement induced $p$-wave trimers
\cite{Levinsen2009,Nishida2010} in polarized gases, in which case the
dimer acts as the impurity investigated here.

\begin{acknowledgments}
  We are grateful to V. Gurarie, M. Lewenstein, and D.~S.~Petrov for
  fruitful discussions.  JL acknowledges support from a Marie Curie
  Intra European grant within the 7th European Community Framework
  Programme.  PM acknowledges funding through MEC project TOQATA and
  ERC Advanced Grant QUAGATUA. CL acknowledges support from the EPSRC
  through Contract No. EP/E053033/1.  JL, PM, and CL wish to thank the
  Aspen Center for Physics, where part of this work was realized. FC
  acknowledges support from R\'egion Ile de France (IFRAF), Institut
  Universitaire de France and ERC (Advanced Grant Ferlodim).
\end{acknowledgments}

\bibliography{pwavepolaronedit,pwave,bibtest2}

\end{document}